\documentstyle[11pt,newpasp,twoside,epsf]{article}
\def\edcomment#1{\iffalse\marginpar{\raggedright\sl#1\/}\else\relax\fi}
\reversemarginpar

\begin{document}
\title{Satellites of Isolated Elliptical Galaxies}
\author{R.M. Smith}
\affil{School of Physics and Astronomy, Cardiff University, P.O. Box 913, Cardiff, CF24 3YB, Wales}
\author{V.J. Martinez}
\affil{Astronomical Observatory. University of Valencia,
P.O. Box 22085,  E-46071 Valencia, Spain}

\begin{abstract}
Using well-defined selection criteria applied to the LEDA galaxy catalogue we
have derived a sample of elliptical galaxies that can be classified as 
isolated. From this we have investigated the neighbourhood of these galaxies
to determine the frequency and radial distribution of faint galaxies around 
them and hence derive an estimate of their surrounding satellite population.
The results are compared and contrasted to the satellite population around
isolated spiral galaxies.
\end{abstract}

\section{Introduction}

The morphology-density relationship for galaxies, first investigated by
Dressler (1980), indicates that ellipticals preferentially occur in regions
of high galactic density, whilst spiral galaxies dominate in the `field'.
Together with 
the properties of `typical' elliptical galaxies, with the domination of the
stellar light by an old, red, population and little evidence of recent
star formation or mergers, this lends support to the hypothesis that such 
galaxies are formed from mergers early on in the evolution of the Universe.
Theoretical simulations also provide clues that elliptical galaxies 
primarily form at early epochs. However, the simulations do suggest that
ellipticals in low density regions form over a much longer timescale and are
therefore likely to show evidence of recent mergers or star formation
activity (e.g. Baugh et al. 1996).
Recent studies of the properties of `isolated' elliptical galaxies have been
inconclusive, with some showing evidence of a young stellar population 
(e.g. Kuntschner et al. 2002) whilst others indicate no recent merger events
(e.g. Silva and Bothun 1998).
A major problem in comparing the results of these 
investigations is the varying definition of an 
isolated galaxy. 
What is required is the
objective selection of a sample of galaxies in low density environments
for direct comparison with both those galaxies in groups and clusters and
also theoretical models. From such a sample we can not only investigate
further the properties of the galaxies themselves but also undertake a detailed
study of their local environment, which simulations suggest 
should provide strong clues as to the formation history of the galaxy.

\section{Sample Selection}

The determination of an objective definition of `isolated' is difficult.
In their study of isolated spiral galaxies, Zaritsky et al. (1993, 1997,
hereafter ZSFW)
used two criteria to select their sample, To be isolated, the magnitude 
difference between a neighbour and the `parent' must be greater than 0.7mag for
galaxies within a projected distance of 1Mpc, or greater than 2.2mag within
500kpc. With a lack of detected 
satellites at distances greater than 500kpc there was {\it a posteriori} 
support for the use of these criteria to select isolated galaxies. We 
therefore use their criteria as a basis for the selection of a sample of
isolated ellipticals.

The absolute magnitude and surface brightness limits reached by the ZSFW
survey were such that it did not enter the dwarf regime. At present there
are major uncertainties in our knowledge of the faint end of the 
Galaxy Luminosity Function, with some evidence of an environmental
dependence (e.g. Driver et al. 1999, Davies et al., this conference).
With the current uncertainty in the LF shape at the faint end, 
we will define a sample
of galaxies that are isolated from other bright galaxies ({$\rm M_B < -17$}).

The availability of catalogues of large numbers of galaxies now makes it viable
to investigate the environmental properties of galaxies and hence select a 
sample of objects that satisfy certain isolation criteria. The LEDA and NED
catalogues are two of the most widely used catalogues. Here we use the LEDA
catalogue which now contains well over a million galaxies from a wide variety of
sources. This catalogue is not complete to any given magnitude but here
we are interested in defining a sample of isolated galaxies that match
certain criteria and do not
investigate the statistics of such a sample. Firstly, the sample of 
ellipticals was defined using the following criteria:-
(i) Redshift less than $10000 kms^{-1}$ -- to ensure the 
sample is almost complete,
(ii) Absolute magnitude $\rm M_B \le 19$ -- to ensure galaxies are `normal',
any satellites found are brighter than the possible 
turn-up in the LF and that the sample is again approximately complete,
(iii) Galactic latitude $\rm >  |25\deg|$
and (iv) Morphological type $\rm t < -4$.

Using these criteria produced a sample of 940 ellipticals, to which the ZSFW
criteria for `isolatedness' were applied. Again the LEDA database was used
to search for neighbours around the central elliptical. In this study no
redshift information was used to select or reject possible isolated
galaxies. The criteria are therefore much stricter than used in 
other studies but ensures that the galaxies are truly isolated. However,
it rules out any statistical study of the frequency of isolated galaxies.
Out of the 940 elliptical galaxies in the sample, 32 satisfied the ZSFW 
criteria. All of these galaxies were cross-checked with the NED catalogue
to ensure they were isolated from bright galaxies in that catalogue.

\section{The Satellite Population}

Previous studies have concentrated on the properties of the central elliptical
yet from the morphology-density relationship it is clear that the environment
must play a very significant part in the evolution of the galaxy. The 
surrounding galaxy population also gives strong clues to the formation history
of the galaxies. Although
the objects selected here are isolated from bright galaxies they may
be surrounded by a halo of fainter, dwarf, galaxies. To investigate this 
possibility we have employed a technique similar to that of Holmberg (1969) and
Lorrimer et al. (1994, hereafter LFSWZ). 
Using the data publicly available from the APM 
plate-scanning
machine, we have selected all galaxies in the field of the parent detected
on the UK and Palomar Schmidt sky survey plates..
Due to the clustering properties of galaxies,
any statistical excess over that of the surrounding area is likely to
be due to surrounding dwarfs. An absolute magnitude limit of
$\rm {M_B}_j = -14.6$ for the surrounding galaxy population (assuming they are
all at the redshift of the parent)
and a redshift limit of $6500 kms^{-1}$ for the primary
was applied  
to ensure that the APM scans 
were reasonably complete for high surface brightness objects (it is 
well-known that at low surface-brightnesses the catalogue is incomplete).
It also ensures that any surrounding galaxies will be dwarfs.
Brighter than $\rm M_B=-16.8$ the
sample of dwarfs is incomplete due to the isolation criteria used to select 
the parent sample. A total of 10 galaxies in the sample of 32 had APM data
suitable for this study.

\begin{figure}
\plotfiddle{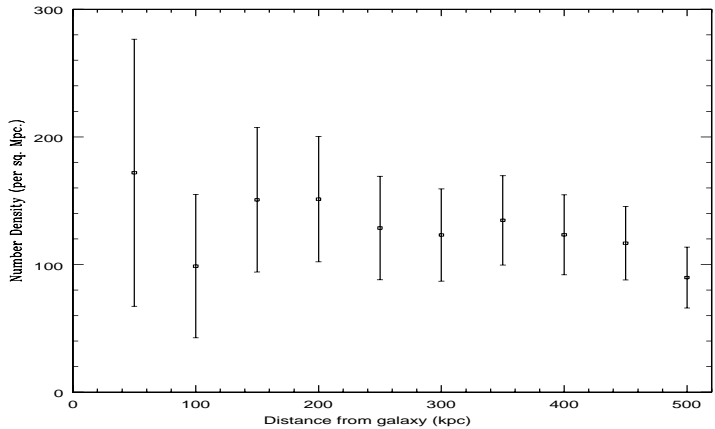}{2.6in}{0}{50}{30}{-150}{-50}

\caption{Radial profile of galaxies surrounding, on average, isolated ellipticals}
\end{figure}
 Due to the large range of redshifts of the parent galaxies the
background population varies widely. In this study we have stacked the profiles
of all 10 of the galaxies to obtain a `mean' profile. This is shown in Figure
1.
It is clear that there is a significant excess of dwarf galaxies within 500kpc.
To obtain an estimate of the dwarf population requires subtraction of the
contaminating background population. With ZFSW finding very few satellites 
beyond 500kpc we use the outer values of the radial profile as an estimate of
the background. Fitting a power law to the
resulting background-subtracted galaxy counts gives a power-law
slope of $\rm -0.6 \pm 0.2$.
This is similar to the slope found for late-type galaxy satellites by LFSWZ 
but less steep than that found by them for early-type galaxies. However, they
found a weak dependence of the slope on the satellite luminosity, with fainter
galaxies having a flatter slope. Extrapolating their 
results to the magnitude limits reached in this study, 
there is good agreement with the
value presented here. They did not find such a dependence for late-type
galaxies.

There are
$\rm 45\pm 15$ dwarfs within 500kpc of the primary down to the limiting 
magnitude of 14.6 and $\rm 19 \pm 6$ with $-16 < M_B < -15$.
Brighter than -16 the number of satellites agrees
with the values of LFSWZ. Comparing the number of faint dwarfs to the
values for brighter satellites implies a steep luminosity function ($\alpha 
\sim -1.8$), in approximate 
agreement with that found for poor clusters (e.g. Driver et al. 1998) and also
the value derived by Morgan et al. (1998) for isolated spirals. LFSWZ also
proposed a steep luminosity funtion around early-type galaxies as an
explanation for the differing clustering length for bright and faint
neighbours. This lends some support to the CDM model of
hierarchical structure formation (e.g. White and Frenk 1991), where there
should be an abundance of small dark matter halos. However, the field 
luminosity function has a slope of $-1.2$ (e.g. Norberg et al. 2002) suggesting
that dwarf galaxies are not the dominant population everywhere.
In addition, comparison
of the radial density profiles for the individual galaxies shows much 
variation, with some galaxies having a very concentrated dwarf population,
some more extended and some have no surrounding dwarfs, suggesting that the
luminosity function is variable. However, the errors are large and thus
a more detailed 
interpretation of the galaxy-to-galaxy variations awaits a deeper CCD survey
of the fields of these galaxies, together with a redshift survey to determine
the satellite population.

\end{document}